\documentclass[prd,twocolumn,superscriptaddress,showpacs,nofootinbib,preprintnumbers]{revtex4}

\usepackage{amsmath}
\usepackage{amsfonts}
\usepackage{graphicx}
\usepackage{dcolumn}

\def\be{\begin{equation}}
\def\ee{\end{equation}}
\def\ba{\begin{eqnarray}}
\def\ea{\end{eqnarray}}
\def\bs{\begin{subequations}}
\def\es{\end{subequations}}

\usepackage{color}

\begin{document}

\title{Chaotic dynamics in preheating after inflation}

\author{Yoshida Jin}
\email{jin@gravity.phys.waseda.ac.jp}
\affiliation{Department of Physics, Waseda University, Okubo 3-4-1,
Shinjuku, Tokyo 169-8555, Japan}
\author{Shinji Tsujikawa}
\email{shinji@nat.gunma-ct.ac.jp}
\affiliation{Department of Physics, Gunma National College of
Technology, Gunma 371-8530, Japan}
\date{\today}

\begin{abstract}
    
We study chaotic dynamics in preheating after inflation in which
an inflaton $\phi$ is coupled to another scalar field $\chi$
through an interaction $(1/2)g^2\phi^2\chi^2$. We first estimate the
size of the quasi-homogeneous field $\chi$ at the beginning of 
reheating for large-field inflaton potentials $V(\phi)=V_0\phi^n$
by evaluating the amplitude of the $\chi$ fluctuations on scales
larger than the Hubble radius at the end of inflation.
Parametric excitations of the
field $\chi$ during preheating can give rise to chaos between two
dynamical scalar fields. For the quartic potential ($n=4$,
$V_0=\lambda/4$) chaos actually occurs for 
$g^2/\lambda <{\cal O}(10)$ in a linear regime 
before which the backreaction of created
particles becomes important. This analysis is supported by several
different criteria for the existence of chaos. For the quadratic
potential ($n=2$) the signature of chaos is not found by the time
at which the backreaction begins to work, similar to the case of
the quartic potential with $g^2/\lambda \gg 1$.
\end{abstract}

\maketitle

\section{Introduction}

Reheating after inflation is an extremely important stage
to generate elementary particles present in current universe.
In the original version of the reheating
scenario which is now called {\it old reheating}, the decay of
an inflaton field is characterized by a perturbative Born
process \cite{oldre}.
However this process is not efficient for the success of the
GUT-scale baryogenesis scenario.
Later it was found that the existence of a nonperturbative
stage--dubbed {\it preheating}-- can lead to an explosive
particle production prior to the Born
decay \cite{TB,KLS1,Boya}.

During preheating scalar particles $\chi$ coupled to the inflaton
$\phi$ are efficiently generated by parametric resonance through
an interaction $(1/2)g^2\phi^2\chi^2$.
The existence of the preheating stage provides several interesting
possibilities such as the GUT-scale baryogenesis \cite{GUT},
nonthermal phase transition \cite{nonther}, the enhancement of
metric perturbations \cite{mpre} and
the formation of primordial black holes \cite{PBH}.
In the chaotic inflationary scenario characterized by the potential
$V(\phi)=V_0\phi^n$, the field perturbations $\delta \chi$ obey
the Mathieu equation (for $n=2$) or the Lame equation (for $n=4$),
which determines the structure of resonance at the linear regime.
When the backreaction of created particles begins to violate
the coherent oscillation of $\phi$, this tends to work to
suppress exponential growth of the field fluctuations.
The system enters a fully nonlinear stage after which the mode-mode
coupling (rescattering) between perturbations is
crucially important \cite{KT,KLS2}.

Typically the contribution of the dynamical background field
$\chi$ is neglected in standard analysis of particle creations in
preheating. This may be justified for a quadratic inflaton 
potential, since large-scale $\chi$ modes are exponentially
suppressed during inflation for the coupling $g$ required for
preheating \cite{Iva}. However the situation is different
for a quartic inflaton potential with the coupling $g$
of order $g^2/\lambda={\cal O}(1)$ \cite{Bruce,Fabio}.
In this case the quasi-homogeneous field $\chi$ can play an
important role for the dynamics of preheating. 
In fact Podolsky
and Starobinsky \cite{Podolsky} pointed out that chaos may occur
for the self-coupling potential $V(\phi)=(1/4) \lambda \phi^4$
when the coupling $g^2/\lambda$ is not too much larger than of
order unity. Since it is not obvious whether chaos actually occurs
or not in this model only by analytic estimations, we shall
perform detailed numerical investigation with/without the
backreaction effect of created particles. 
We shall estimate the size of the field $\chi$ at 
the beginning of reheating by evaluating the amplitude of 
the $\chi$ fluctuations for the modes larger than the
Hubble radius.
To judge the existence of chaos in preheating
we will adopt several different methods-- such as the 
Toda-Brumer test \cite{Toda,Brumer}, Lyapunov exponents \cite{Lya}
and a fractal map \cite{Barrow}.

It was already found that chaos appears for hybrid-type inflation
models \cite{EM,Bellido,Bastero,Zibin}
(see also Refs.~\cite{Cornish,BT,Joras}).
Hybrid inflation is a rather special model in a sense that the 
symmetry breaking
field automatically grows by tachyonic instability even if it is 
suppressed
during inflation. The necessary condition for chaos is that there 
exist at least two
dynamical fields and neither of them is too much smaller than another 
field.
The hybrid model satisfies this condition, since two fields can have 
frequencies
which are the same order after symmetry breaking.
The presence of mixing terms between two fields
leads to a new instability of perturbations in addition to 
tachyonic/resonance
instabilities \cite{Bastero}.
This new type of instability is clearly associated with the presence 
of chaos.

In this work we shall investigate the existence of chaos for
large-field potentials $V(\phi)=V_0\phi^n$ with an interaction 
$(1/2)g^2\phi^2\chi^2$.
We estimate the variance of large-scale modes in $\chi$ at 
the end of inflation, which is relevant to the initial condition of 
the 
quasi-homogeneous field $\chi$ for preheating. 
The field $\chi$ is amplified by parametric
resonance, which can give rise to chaos after $\chi$ grows to
satisfy the Toda-Brumer condition.
We shall numerically solve background equations
together with perturbed equations for both quartic ($n=4$) and 
quadratic ($n=2$) potentials.
Our main interest is to find the signature of chaos and the parameter 
range of
the coupling $g$ in which chaos can be seen before the backreaction 
effect of
created particles becomes important.
Since chaos can alter the standard picture of preheating by 
parametric resonance,
it is of interest to clarify the situation in which chaos appears.

Recent observations suggest that the quartic potential
$V(\phi)=(1/4)\lambda \phi^4$ is under an observational pressure,
while the quadratic potential $V(\phi)=(1/2)m^2\phi^2$ is allowed
\cite{obser}. This depends on the number of $e$-folds $N$ before
the end of inflation at which observable perturbations are
generated. In the case of quartic potential this corresponds to $N
\sim 64$ by assuming instant transitions between several
cosmological epochs \cite{efolds}. The likelihood analysis
including WMAP and SDSS datasets shows that the quartic potential
is marginally allowed by using $N=64$ \cite{Tegmark}. Therefore it
is premature to rule out this model completely from current
observations. The quartic potential corresponds to a system in
which the background equations can be reduced to the ones in
Minkowski spacetime by introducing conformal variables. This has
an advantage for the investigation of chaotic dynamics during
preheating. As we see later, the quartic potential exhibits a
stronger chaos compared to the one for the quadratic potential.

\section{The field variance for long wavelength modes}

In this section we shall estimate the field variance
for long wavelength modes of the field
$\chi$ coupled to the inflaton $\phi$ through an
interaction $(1/2)g^2\phi^2\chi^2$.
The effective potential in our system is
\begin{eqnarray}
V(\phi, \chi)=V_0\phi^n+\frac12 g^2\phi^2\chi^2\,.
\label{effpo}
\end{eqnarray}
We are mainly interested in two inflaton potentials:
(i) the quadratic one ($n=2$) and (ii) the quartic one ($n=4$).
In this work we do not implement nonminimal 
couplings \cite{nonmini}
between the field $\chi$ and the scalar curvature $R$.

In a flat Friedmann-Lemaitre-Robertson-Walker (FLRW) background
with a scale factor $a$, the background equations
for the system (\ref{effpo}) are
\begin{eqnarray}
\label{back2}
& &\ddot{\phi}+3H\dot{\phi}+V_{\phi}=0\,,\\
\label{back3}
& &\ddot{\chi}+3H\dot{\chi}+V_{\chi}=0\,, \\
\label{back}
& & H^2 \equiv \left(\frac{\dot{a}}{a}\right)^2
=\frac{\kappa^2}{3}
\left[\frac12 \dot{\phi}^2+ \frac12 \dot{\chi}^2+
V(\phi,\chi)\right]\,,
\end{eqnarray}
where $V_\phi=nV_0\phi^{n-1}+g^2\chi^2\phi$,
$V_\chi=g^2\phi^2\chi$ and
$\kappa^2 = 8\pi/m_p^2$ with $m_p$
being the Planck mass.

We define the number of $e$-folds before the end of inflation, as
\begin{eqnarray}
N \equiv {\rm ln}\,(a_f/a(t))\,,
\end{eqnarray}
where $a_f$ is the scale factor at the end of inflation.
Employing the slow-roll approximation, $|\ddot{\phi}| \ll 
|3H\dot{\phi}|$
and $\dot{\phi}^2 \ll V(\phi)$, we easily find that
${\rm d}\phi/{\rm d}N=n/(\kappa^2\phi)$.
Here we neglect the contribution coming from the $\chi$-dependent 
terms.
Integrating this equation, we obtain
\begin{eqnarray}
\phi^2-\phi_f^2=\frac{2n}{\kappa^2}N\,.
\end{eqnarray}
The field value at the end of inflation ($\phi_f$) is determined
by setting the slow-roll parameter, $\epsilon_{s} \equiv (1/2\kappa^2)
(V_\phi/V)^2$, is unity.
This gives $\phi_f/m_p=n/(4\sqrt{\pi})$,
thereby leading to
\begin{eqnarray}
\label{phislow}
\phi^2=\frac{n}{4\pi} \left(N+\frac{n}{4}\right)
m_p^2\,.
\end{eqnarray}

Let us consider a perturbation $\delta \chi$ in the field 
$\chi$. Then the momentum-space first-order perturbed 
equation is given by 
\begin{eqnarray}
\label{dchikeq}
\delta \ddot{\chi}_{k}+3H\delta \dot{\chi}_{k}+
\left(\frac{k^2}{a^2}+g^2\phi^2 \right)\delta \chi_{k}=0\,,
\end{eqnarray}
where $k$ is a comoving wavenumber.
Note that we neglected the backreaction of 
gravitational perturbations.
When the effective mass of the field $\chi$ is larger than 
of order the Hubble rate, i.e., $g^2\phi^2>(3H/2)^2$,  
the evolution of super-Hubble perturbations is characterized by 
underdamped oscillations with the dependence \cite{Iva} 
\begin{eqnarray}
\label{chi1}
\delta \chi_{k} \propto a^{-3/2}\,.
\end{eqnarray}
Hence large-scale perturbations are exponentially suppressed 
during inflation. 
Meanwhile when $g^2\phi^2<(3H/2)^2$ super-Hubble 
perturbations evolve as
\begin{eqnarray}
\label{chi2}
\delta \chi_{k} \propto \exp \left[ -\left(
\frac32 H-\sqrt{\frac94 H^2-g^2\phi^2} 
\right){\rm d}t \right]\,.
\end{eqnarray}
This shows that in the massless limit ($g^2\phi^2 \ll H^2$)
the amplitude of $\delta \chi_{k}$ decreases very slowly.
In what follows we shall consider the quadratic and quartic 
models separately.

\subsection{Quadratic model}

For the inflation potential $V(\phi)=\frac12m^2\phi^2$, 
the coupling $g$ is required to be greater than of order 
$10^{-4}$ in order for preheating to occur \cite{KLS2}.
In this case the resonance parameter, $q \equiv g^2\phi^2/(4m^2)$,
is much larger than 1 at the beginning of preheating \cite{KLS2}.
Since $H \sim m$ at the end of inflation, 
the existence of the preheating stage demands the condition
$g^2\phi^2 \gg H^2$ during inflation.
Hence the evolution of large-scale $\chi$ fluctuations
is characterized by Eq.~(\ref{chi1}).

Let us consider the modes which are outside the Hubble radius at 
the end of inflation ($0<k<k_f=a_{f}H_{f})$.
Since these modes are effectively massive with slowly changing mass, 
they can be treated as an adiabatic state characterized by 
$\delta \chi_{k}=1/(a^{3/2}\sqrt{2\omega_{k}})$ where 
$\omega_k^2 \simeq g^2\phi^2$.
Hence the amplitude $|\delta \chi_{k}|^2$ at the end of 
inflation is estimated as 
\begin{eqnarray}
\label{chiend}
|\delta \chi_{k} (t_{f})|^2=\frac{1}{2a_{f}^3g\phi_f}\,.
\end{eqnarray}

Then we can obtain the variance of the fluctuation 
$\delta \chi_{k}$ for $k<k_f$:
\begin{eqnarray}
\langle \delta \chi^2_{k} (t_{f}) \rangle_{k<k_{f}}
&=&
\frac{1}{2\pi^2} \int_{0}^{k_f}
k^2 |\delta \chi_{k} (t_{f})|^2 {\rm d} k \nonumber \\
&=& \frac{\phi_{f}^2}{9\pi g} 
\sqrt{\frac{4\pi}{3}}
\left(\frac{m}{m_{p}}\right)^3\nonumber \\
\label{dchik}
&=& \frac{1}{9\pi \sqrt{12\pi} g}
\left(\frac{m}{m_{p}}\right)^3 m_{p}^2\,,
\end{eqnarray}
where we used $\phi_{f}=m_{p}/(2\sqrt{\pi})$
in addition to the slow-roll approximation
$H^2 \simeq 4\pi m^2\phi^2/3m_p^2$.

\begin{figure}
\begin{center}
\includegraphics[height=3.2in,width=3.4in]{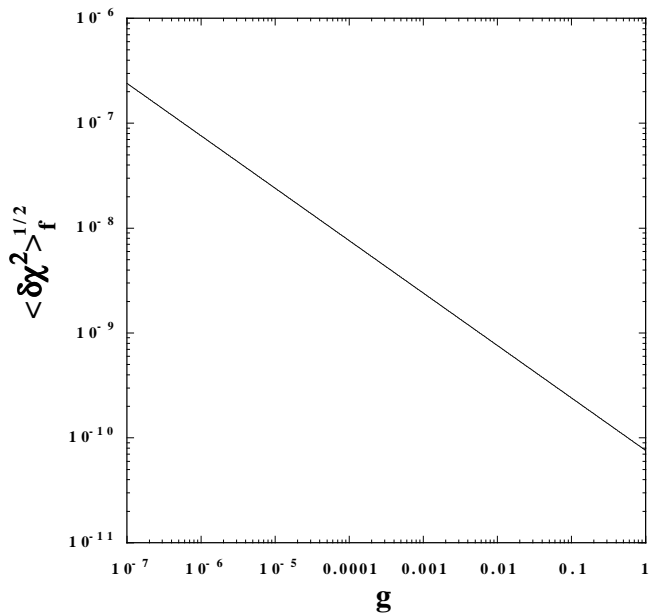}
\caption{The amplitude of the  variance 
$\langle \delta \chi^2_{k}
(t_{f}) \rangle_{k<k_f}^{1/2}$ at the end of inflation
in terms of the function of the coupling $g$
for the quadratic potential ($n=2$).
The variance gets smaller for larger $g$.
}
\label{n2}
\end{center}
\end{figure}

For the quadratic potential ($n=2$) the inflaton mass is
constrained to be $m \simeq 10^{-6}m_p$ from
the COBE normalization \cite{LL}.
Then Eq.~(\ref{dchik}) shows that the variance is 
the function of $g$ only.
In Fig.~\ref{n2} we plot $\sqrt{\langle \delta \chi^2_{k}
(t_{f}) \rangle_{k<k_f}}$ as a function of $g$.
It can be regarded as a quasi-homogeneous mode 
at the beginning of preheating.
This at least measures the minimum amplitude  
of the homogeneous $\chi$ field.

\subsection{Quartic model}

For the quartic potential $V(\phi)=(1/4)\lambda\phi^4$, 
it is known that preheating occurs even when the coupling 
$g$ is in the range $g^2\phi^2 \lesssim H^2$ \cite{Kaiser,GKLS}.
When $g^2/\lambda={\cal O}(1)$, for example, the background
dynamics transits from the ``massless regime'' 
[$g^2\phi^2<(3H/2)^2$] to the ``massive regime'' 
[$g^2\phi^2>(3H/2)^2$] during inflation \cite{Bruce,Fabio,Zibin2}.  
The critical number of $e$-folds, $N_c$,  in which the 
the evolution of the perturbation $\delta \chi_{k}$ transits
from Eq.~(\ref{chi2}) to Eq.~(\ref{chi1}) is determined by the 
condition 
$9H_c^2/4=g^2\phi_{c}^2$, which gives \cite{Zibin2}
\begin{eqnarray}
N_{c}=\frac23 \frac{g^2}{\lambda}
=\ln \left(\frac{a_{f}}{a_{c}}\right) 
\simeq \ln \left(\frac{k_{f}}{k_{c}}\right)\,.
\label{NC}
\end{eqnarray}
We note that we used the slow-roll condition 
\begin{eqnarray}
\label{Quarback}
H^2 \simeq \frac{2\pi \lambda \phi^4}{3m_{p}^2},\quad
\phi^2 \simeq \frac{N}{\pi}m_p^2\,.
\end{eqnarray}

The modes which are inside the Hubble radius at transition 
time $t=t_{c}$
(but larger than the Hubble radius at the end of inflation)
evolves as effective massive fields for $t>t_{c}$.
Then for $k_{c}<k<k_{f}$ the amplitude 
$|\delta \chi_{k}|^2$ at the end of 
inflation is given by Eq.~(\ref{chiend}), 
which gives the variance
\begin{eqnarray}
\langle \delta \chi^2_{k} (t_{f}) 
\rangle_{k_{c}<k<k_{f}}
&=&
\frac{1}{2\pi^2} \int_{k_{c}}^{k_f}
k^2 |\delta \chi_{k} (t_{f})|^2 {\rm d} k \nonumber \\
\label{delchikd}
&=& \frac{\lambda}{18\pi^3}
\sqrt{\frac{2\lambda}{3g^2}}
(1-e^{-2g^2/\lambda})m_p^2\,, \nonumber \\
\end{eqnarray}
where we used the relation $k_{c}=k_{f}\exp (-2g^2/3\lambda)$
coming from Eq.~(\ref{NC}).

For $k<k_{c}$ the modes exit the Hubble radius before
the transition time $t=t_c$.
At the Hubble radius crossing ($t=t_k$) the amplitude of the 
perturbation 
$\delta \chi_{k}$ is given by 
\begin{eqnarray}
|\delta \chi_k(t_{k})|^2=\frac{H^2(t_{k})}{2k^3}\,,
\end{eqnarray}
which comes from the quantization of a standard massless
scalar field \cite{LL}.
Since $\delta\chi_k$ evolve as Eq.~(\ref{chi2}) for
$t_{k}<t<t_{c}$, we find that the amplitude of the 
perturbations at $t=t_c$ is given by \cite{Zibin2}
\begin{eqnarray}
|\delta \chi_{k} (t_{f})|^2=
\frac{H^2(t_{k})}{2k^3}e^{-3F(N_{k})}\,,
\end{eqnarray}
where
\begin{eqnarray}
F(N_k)&=&N_{k}-N_{c}-\sqrt{N_{k}(N_{k}-N_{c})} 
\nonumber \\
& &+N_{c} \ln \left( \frac{\sqrt{N_{k}}+
\sqrt{N_{k}-N_{c}}}{\sqrt{N_{c}}}\right)\,.
\end{eqnarray}
Here $N_{k}$ is the number of $e$-folds at $t=t_{k}$.
The perturbations evolve as Eq.~(\ref{chi2}) 
for $t_{c}<t<t_{f}$.
Using Eq.~(\ref{NC}) we obtain 
the following amplitude at the end of inflation:
\begin{eqnarray}
|\delta \chi^2_{k} (t_{f})|=
\frac{H^2(t_{k})}{2k^3}e^{-3F(N_{k})-2g^2/\lambda}\,.
\end{eqnarray}
Then the variance of perturbations $\delta \chi_{k}$
for the modes $k<k_{c}$ is given by 
\begin{eqnarray}
\langle \delta \chi_{k}^2 (t_{f}) \rangle_{k<k_{c}}
&=&
\frac{1}{2\pi^2} \int_{k_i}^{k_c}
|\delta \chi_{k} (t_{f})|^2 k^3 {\rm d}(\ln k) \nonumber \\
&=& \frac{1}{4\pi^2}
\int_{N_{c}}^{N_{i}} {\rm d}N_k H^2(t_{k})
e^{-3F(N_{k})-2g^2/\lambda}\,, \nonumber \\
&=& \frac{\lambda m_p^2}{6\pi^3}
\int_{N_{c}}^{N_{i}} {\rm d}N_k
N_k^2 e^{-3F(N_{k})-2g^2/\lambda}\,,
\end{eqnarray}
where we used Eq.~(\ref{Quarback}) and 
${\rm d}(\ln k) \simeq {\rm d}(\ln a)=-{\rm d}N_k$.
Note that $k_{i}$ is the minimum wavenumber relevant for 
the maximum scale of cosmological perturbations.
In order to obtain 
$\langle \delta \chi^2_{k} (t_{f}) \rangle_{k<k_{c}}$, 
it is necessary to solve the following differential equation:
\begin{eqnarray}
\frac{{\rm d}}{{\rm d}N_{k}}
\langle \delta \chi^2_{k} (t_{f}) \rangle_{k<k_{c}}
=\frac{\lambda m_p^2}{6\pi^3}
N_k^2 e^{-3F(N_{k})-2g^2/\lambda},
\end{eqnarray}
which should be integrated from $N_{c}=2g^2/3\lambda$
to $N_{i}$. Note that $N_{i}$ roughly corresponds to the
total number of $e$-folds during inflation. 
At least we require the condition $N_{i}>60$.
We shall choose $N_{i}=60, 100, 1000$
in order to see the sensitivity for the change of this number.

\begin{figure}
\begin{center}
\includegraphics[height=3.2in,width=3.4in]{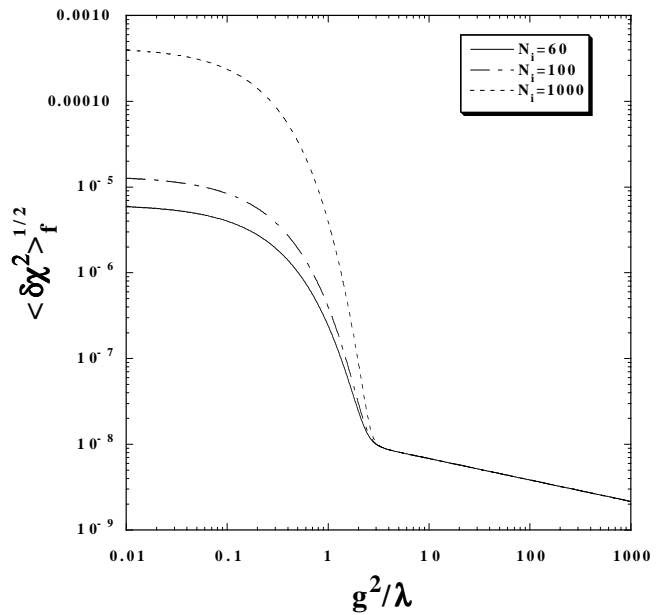}
\caption{The amplitude of the  variance 
$\langle \delta \chi^2_{k}
(t_{f}) \rangle_{k<k_f}^{1/2}$ at the end of inflation
in terms of the function of the coupling $g$
for the quartic potential ($n=4$).
The variance is dominated by the modes $k_{c}<k<k_{f}$
for $g^2/\lambda \gg 1$, whereas the dominant contribution 
comes from the modes $k<k_{c}$ for 
$g^2/\lambda<{\cal O}(1)$.
}
\label{n4}
\end{center}
\end{figure}

Finally the variance for the modes $k<k_{f}$ is given by the 
following sum:
\begin{eqnarray}
\langle \delta \chi^2_{k} (t_{f}) \rangle_{k<k_f}
=\langle \delta \chi^2_{k} (t_{f}) \rangle_{k_{c}<k<k_{f}}
+\langle \delta \chi^2_{k} (t_{f}) \rangle_{k<k_c}\,.
\end{eqnarray}
When $g^2/\lambda \lesssim {\cal O}(1)$ one has
$N_{c} \lesssim {\cal O}(1)$. This shows that most of 
the contribution to the variance 
$\langle \delta \chi^2_{k} (t_{f}) \rangle_{k<k_f}$ 
comes from the modes $k<k_{c}$.
Meanwhile when $g^2/\lambda \gg 1$, i.e., $N_{c} \gg 1$,
the modes $k_{c}<k<k_{f}$ dominate the total variance.
In Fig.~\ref{n4} we plot $\sqrt{\langle \delta \chi^2_{k} (t_{f}) 
\rangle_{k<k_f}}$ as a function of $g^2/\lambda$
for $N_{i}=60, 100, 1000$ and $\lambda=10^{-13}$.
We find that the variance does not depend on the 
values $N_{i}$ for $g^2/\lambda \gtrsim 3$, which reflects the fact that 
Eq.~(\ref{delchikd}) is independent of $N_{i}$.
The difference appears for $g^2/\lambda \lesssim 3$,
because of the fact that 
$\langle \delta \chi^2_{k} (t_{f}) \rangle_{k<k_c}$
is dependent on the values $N_{i}$.
We shall use the values $\langle \delta \chi^2_{k} (t_{f}) 
\rangle_{k<k_f}$
obtained for $N_{i}=60$ as a minimum initial condition 
of the quasi-homogeneous $\chi$ field
at the beginning of preheating.

We note that Podolsky and Starobinsky \cite{Podolsky} 
estimated the size of the quasi-homogeneous $\chi$ field
for $g^2/\lambda={\cal O}(1)$
by using a Fokker Planck equation \cite{Sta}.
This equation is valid when the mass of the 
field $\chi$ is smaller than of order the Hubble 
rate \cite{NNS,Hosoya89,Salopek}, 
which corresponds to $g^2/\lambda  \lesssim {\cal O}(1)$.
We checked that our estimation of the variance shows good agreement 
with the one based on the Fokker Planck approach for 
$g^2/\lambda  \lesssim {\cal O}(1)$.
In the parameter regime $g^2/\lambda \gg 1$ 
the Fokker Planck equation is no longer valid.
Hence we need to use our estimation given above 
in order to know the size of the quasi-homogeneous $\chi$ field
at the end of inflation.

\section{Basic properties of preheating and chaos}

\subsection{Preheating and the role of
the quasi-homogeneous field $\chi$}

In the presence of the coupling $(1/2)g^2\phi^2\chi^2$, the
coherent oscillation of the inflaton leads to the excitation of
the field $\chi$ during preheating through the resonance term
$g^2\phi^2 \chi$ in Eq.\,(\ref{back3}). For the quadratic potential
($n=2$) parametric resonance is efficient when the condition, 
$q= g^2\phi^2/(4m^2) \gg 1$, is satisfied. To be
more precise the field $\chi$ grows for $g \gtrsim 3.0 \times
10^{-4}$ by overcoming the friction due to cosmic expansion
\cite{KLS2}. In this model the growth of $\chi$ ends when the
system enters a narrow resonance regime ($q \lesssim 1$) or the
backreaction effect of created particles breaks the coherent
oscillation of the field $\phi$. For the quartic potential ($n=4$)
resonance bands exist for the parameter space characterized by
$n(2n-1)<g^2/\lambda<n(2n+1)$, where $n=1, 2, 3,...$
\cite{Kaiser,GKLS}. The center of the band, $g^2/\lambda=2n^2$,
corresponds to the largest Floquet index ($\mu_{\rm max} \simeq
0.28$).

If the field $\chi$ is strongly suppressed during inflation,
this does not contribute to the background dynamics
even if it is amplified during
reheating\footnote{The amplification of $\chi$ is limited by
the backreaction effect of created particles.}.
As we see in Figs.~\ref{n2} and \ref{n4}, it is expected 
that $\chi$ does not dynamically 
become important during preheating for larger $g$.
However, for the quartic potential, parametric resonance 
takes place even for $g^2/\lambda={\cal O}(1)$, 
in which case the quasi-homogeneous $\chi$ field 
does not suffer from strong
suppression during inflation.
Then if the quasi-homogeneous field $\chi$ is amplified during 
preheating,
this can contribute to the background dynamics.

In our model the equations for field perturbations in Fourier space
may be written as \cite{Podolsky}
\begin{eqnarray}
\label{dphi}
& &\delta \ddot{\phi}_k+3H\delta\dot{\phi}_k+
\left[\frac{k^2}{a^2}+n(n-1)V_0\phi^{n-2}+g^2\chi^2\right]
\delta \phi_k \nonumber \\
& & =-2g^2\phi\chi \delta \chi_k   \,, \\
& &\delta \ddot{\chi}_k+3H\delta\dot{\chi}_k+
\left(\frac{k^2}{a^2}+g^2\phi^2\right)\delta \chi_k
\nonumber \\
& & =-2g^2\phi\chi \delta \phi_k\,,
\label{dchi}
\end{eqnarray}
where $k$ is a comoving wavenumber.
This corresponds to the equations in which
the contributions from metric perturbations
are dropped (see Ref.~\cite{mpre}).
Since metric perturbations are enhanced only when field
fluctuations grow sufficiently, it is a good approximation to
neglect them except for a nonlinear stage of preheating.

The terms on the r.h.s. of Eqs.~(\ref{dphi}) and (\ref{dchi}) are
not usually taken into account in standard analysis of preheating
\cite{TB,KLS1,KLS2}, since the field $\chi$ was supposed to be
dynamically unimportant. However this can give considerable
contributions to the dynamics of field perturbations provided that 
the quasi-homogeneous field $\chi$ is not strongly suppressed 
during inflation.
In fact these terms lead to a mixing between the perturbations of
two fields, whose behavior is absent in standard analysis of
preheating unless rescattering effects are taken into account at a
nonlinear stage.

We start integrating background and perturbation
equations from the end of inflation.
We adopt the Bunch-Davies vacuum state for the initial
condition of perturbations for the modes inside the
Hubble radius.
The total variances of the fields $\varphi=\phi, \chi$
integrated in terms of $k$ are
\begin{eqnarray}
\langle \delta \varphi^2 \rangle=
\frac{1}{2\pi^2} \int k^2
|\delta \varphi_k|^2 {\rm d}k\,.
\end{eqnarray}
We implement the variances $\langle \delta \phi^2 \rangle$
and $\langle \delta \chi^2 \rangle$
for both background and perturbation equations
as a Hartree approximation \cite{KLS2}.
We note that this is for estimating
the time at which backreaction effects become important.
After the system enters a fully nonlinear stage,
one can not trust the analysis using the Hartree approximation.
Our interest is to find a signature of chaos
before the backreaction sets in.

\subsection{The condition for chaos}

In this subsection we review several conditions for the
existence of chaos and apply them to our effective
potential (\ref{effpo}).
Let us consider the first-order differential equations
\begin{eqnarray}
\label{dyna}
\dot{x}_i=F_i(x_j)\,,
\end{eqnarray}
and their linearized equations,
\begin{eqnarray}
\delta \dot{x}_i
=\frac{\partial F_i}
{\partial x_j}\, \delta x_j\,,
\label{eq:linearized}
\end{eqnarray}
where $\delta x_i$ is the perturbation vector connecting
two nearby trajectories
and $\partial F_i/\partial x_j$ is the Jacobian matrix
of $F_i(x_j)$.

The scalar-field equations (\ref{back2}) and (\ref{back3}) are 
expressed
by the form (\ref{dyna}) by setting $x_1=\phi$, $x_2=\dot{\phi}$,
$x_3=\chi$ and $x_4=\dot{\chi}$.
Then one can evaluate the Jacobian matrix
$\partial F_i/\partial x_j$ and its eigenvalues $\mu$
for a general system characterized by an effective potential 
$V=V(\phi, \chi)$.
Note that we do not account for the linearized
equation for $H$, since metric perturbations are neglected
in our analysis.
The eigenvalues of the matrix
$\partial F_i/\partial x_j$ are given by
\begin{eqnarray}
\label{mu}
\mu=\frac12 \left[-3H \pm \sqrt{9H^2+4\gamma}
\right]\,,
\end{eqnarray}
where
\begin{eqnarray}
\label{gam}
\gamma&=&\frac12 \biggl[ -(V_{\phi\phi}+V_{\chi\chi})
\nonumber \\
& & \pm
\sqrt{(V_{\phi\phi}+V_{\chi \chi})^2-4(V_{\phi \phi}
V_{\chi \chi}-V_{\phi \chi}^2)}\biggr]\,.
\end{eqnarray}

The necessary condition for the existence of chaos is that one of the
eigenvalues is at least positive. In an expanding background ($H>0$)
this corresponds to the condition $\gamma>0$ from Eq.~(\ref{mu}).
Since we are now considering a situation in which both effective 
masses of
$\phi$ and $\chi$ are positive ($V_{\phi\phi}>0$, $V_{\chi\chi}>0$),
$\gamma$ can take a positive value when
\begin{eqnarray}
V_{\phi \phi} V_{\chi \chi}-V_{\phi \chi}^2<0\,.
\end{eqnarray}
This is so-called the Toda-Brumer test \cite{Toda,Brumer}
that is used to judge the existence of chaos.
For our effective potential (\ref{effpo})
this translates into the condition
\begin{eqnarray}
\label{chicon}
\chi^2>\frac{n(n-1)V_0}{3g^2}\phi^{n-2}\,.
\end{eqnarray}

When $n=2$ and $V_0=m^2/2$, this corresponds to
\begin{eqnarray}
\label{chin=2}
\chi>\frac{m}{\sqrt{3}g}\,,
\end{eqnarray}
whereas for $n=4$ and $V_0=\lambda/4$ we get
\begin{eqnarray}
\chi>\sqrt{\frac{\lambda}{g^2}}\,\phi\,.
\label{chin=4}
\end{eqnarray}
For the quadratic potential the initial value of $\chi$ is
$10^{-4}\sqrt{g}$ times smaller
than the value which leads to chaotic instability,
see Eqs.~(\ref{dchik}) and (\ref{chin=2}).
Therefore the system is expected to enter a chaotic phase after
the field is amplified more than $10^4/\sqrt{g}$ times.
For the quartic potential the condition for chaos is not so severe
compared to the quadratic potential, since the term on the
r.h.s. of Eq.~(\ref{chin=4}) decreases with time.

It is worth commenting on the difference about the instabilities of
chaos and parametric resonance.
Although the field $\chi$ exhibits an exponential growth by
resonance, this is different from the chaotic instability
in which the evolution of the system is very sensitive to slight
change of initial conditions.
In fact none of the
eigenvalues of the Jacobi matrix is positive in the regime
where the condition (\ref{chicon}) is not satisfied.
This means that chaos is absent in the region
$\chi^2<n(n-1)V_0/(3g^2)\phi^{n-2}$ even if the
field shows an exponential growth by parametric resonance.

Since the Toda-Brumer test is not a sufficient condition for the
existence of chaos, we shall use other criteria as well such as
Lyapunov exponents \cite{Lya}.
The Lyapunov exponents measure the logarithm of
of the expansion of a small volume in a $N$-dimensional
phase space. In this case the system possesses $N$
Lyapunov exponents. Chaos is accompanied by an
increase of the size of the $N$-volume at least in one direction and
a maximal Lyapunov exponent $h$ characterizes
the signature of chaos.
When $h$ approaches a positive constant asymptotically,
this shows the existence of chaos since the initial displacement of
the $N$-volume grows exponentially.
If $h$ approaches to 0 asymptotically, this means the
absence of chaos since the orbits are periodic or quasi-periodic.
Strictly speaking Lyapunov exponents are defined in the
limit $t \to \infty$.
Nevertheless one can check the existence of chaos by
investigating the behavior of the system for sufficiently
large values of $t$.
In fact, as we see later, the maximal Lyapunov exponent
begins to grow toward a constant value when chaos appears.
Although the backreaction effect of created particles can alter the
background dynamics, it is possible to see the signature of chaos
before the backreaction sets in.

In addition to the above two criteria, there exists another criterion
for the existence of chaos-- which is so called
a fractal map \cite{Barrow}.
This strategy is useful because of gauge independence which
comes from using a topological character.
In the next section we shall also use this criterion
to confirm the presence of chaos in addition to other methods.

\section{Chaotic dynamics for the quartic potential}

\begin{figure}
\begin{center}
\includegraphics[height=3.3in,width=3.3in]{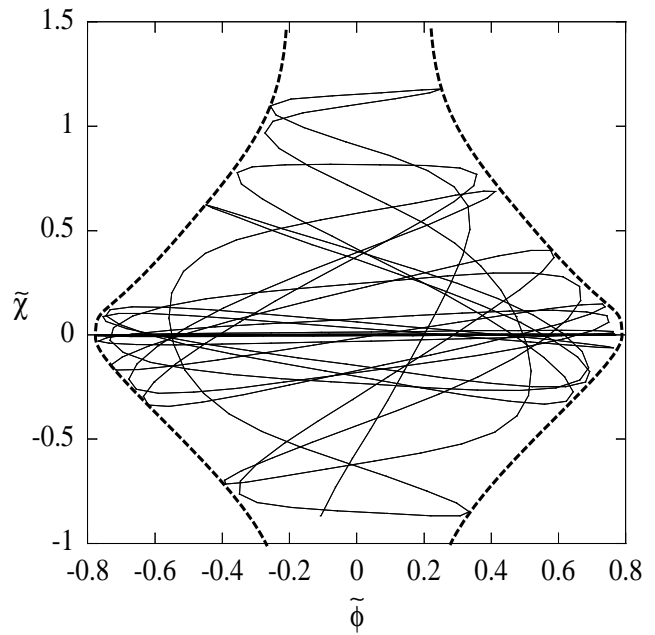}
\caption{A phase-space trajectory in the ($\phi$, $\chi$) plane for
$g^2/\lambda=2$. The dotted curve corresponds to the boundary
given in Eq.~(\ref{bound}).
}
\label{pspace}
\end{center}
\end{figure}

For the quartic potential ($n=4$) the system is effectively
reduced to a Hamiltonian system in Minkowski spacetime
by introducing conformal variables: $\tilde{\phi}\equiv a\phi$
$\tilde{\chi}\equiv a\chi$ and $\eta \equiv \int a^{-1} dt$.
Using the fact that $a \propto \eta$ and $a'/a,\ a''/a \to 0$
during reheating in this model, the background equations
(\ref{back2}), (\ref{back3}) and (\ref{back}) can be written as
\begin{eqnarray}
& & \tilde{\phi}''+\lambda\tilde{\phi}^3
+g^2\tilde{\chi}^2\tilde{\phi}=0\,, \\
& & \tilde{\chi}''+g^2\tilde{\phi}^2\tilde{\chi}=0, \\
a'{}^2 &=& \frac{8\pi}{3m_p^2}\left(
 \frac{1}{2} \tilde{\phi}'{}^2
+\frac{1}{2} \tilde{\chi}'{}^2
+\frac{\lambda}{4} \tilde{\phi}^4
+\frac{g^2}{2}\tilde{\phi}^2\tilde{\chi}^2\right) \nonumber \\
&\equiv& \frac{8\pi}{3m_p^2}E={\rm const}\,,
\end{eqnarray}
where a prime denotes the derivative with respect to conformal
time $\eta$.  The trajectories of the fields are bounded by
\begin{eqnarray}
\label{bound}
\frac{\lambda}{4} \tilde{\phi}^4
+\frac{g^2}{2}\tilde{\phi}^2\tilde{\chi}^2
\le E\,.
\end{eqnarray}
This boundary is plotted as a dotted curve in Fig.~\ref{pspace}.

{}From the Toda-Brumer test (\ref{chin=4}),
we can expect that chaos occurs when $\chi$ becomes
comparable to $\phi$ for $g^2/\lambda={\cal O}(1)$.
When $1<g^2/\lambda<3$, corresponding to the first
resonance band \cite{Kaiser,GKLS},
the variance of the field $\chi$ is larger than 
$\chi_{f} \sim 10^{-8} m_p$ right after the end 
of inflation from Fig.~\ref{n4}.
We require the parametric excitation of $\chi$ 
to give rise to chaos, since
$\chi$ is much smaller than $\phi$ at the beginning of reheating.
Therefore the coupling $g$ needs to lie inside of
the resonance band for the existence of chaos.
In Fig.~\ref{pspace} we show an example of
the background trajectory for $g^2/\lambda=2$.
We find that two fields evolve chaotically  in the phase-space of
the ($\phi, \chi$) plane by choosing several different initial 
conditions.

\begin{figure}
\begin{center}
\includegraphics[height=3.0in,width=3.3in]{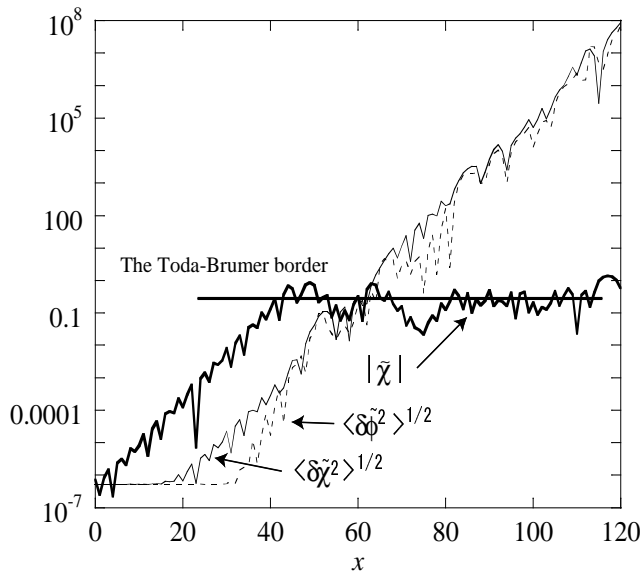}
\caption{Evolution of $\tilde{\chi}$, $\langle \delta \tilde{\chi}^2
\rangle^{1/2}$ and $\langle \delta \tilde{\phi}^2\rangle^{1/2}$
(normalized by $m_p$) as a function of $x \equiv
\sqrt{\lambda} \phi_I \eta$ for $g^2/\lambda=2$
when the backreaction effect of created particles is
neglected. The horizontal line shows the border of the Toda-Brumer 
test.
The perturbations in $\phi$ and $\chi$ exhibit instabilities
associated with chaos once the field $\chi$
is sufficiently amplified.
}
\label{chidelchi}
\end{center}
\end{figure}

When $g^2/\lambda=2$ the Toda-Brumer test gives the condition
$\tilde{\chi}/m_p \gtrsim 0.5$ for the existence of chaos, which
corresponds to the time $x \equiv \sqrt{\lambda}\phi_I\eta \gtrsim 45$
(see Fig.~\ref{chidelchi}).
We find that the fluctuations $\langle \delta \tilde{\chi}^2 \rangle$
and $\langle \delta \tilde{\phi}^2 \rangle$ exhibit rapid increase
with a similar growth rate after the field $\chi$ satisfies
the Toda-Brumer test.
It is expected that this is associated with the presence of chaos
rather than parametric excitation of the $\chi$ fluctuation.
The quasi-homogeneous field $\chi$ is amplified by parametric 
resonance
for $x \lesssim 45$ but stops growing after that.
This comes from the fact that the resonance does not occur
once the homogeneous oscillation of $\phi$ is broken by the growth
of $\chi$. The $g^2\phi^2$ term on the
l.h.s. of Eq.~(\ref{dchi}) also becomes ineffective at this stage, but
the presence of the mixing term on the r.h.s. leads to
a new type of instability associated with chaos.
The rapid growth of $\langle \delta \tilde{\phi}^2 \rangle$
seen in Fig.~\ref{chidelchi} also comes from
the mixing term on the r.h.s. of Eq.~(\ref{dphi})
rather than from the parametric excitation of sub-Hubble modes with
$3/2<k^2/(\lambda \phi_I^2)<\sqrt{3}$ \cite{Kaiser,GKLS}.

\begin{figure}
\begin{center}
\includegraphics[height=3.3in,width=3.5in]{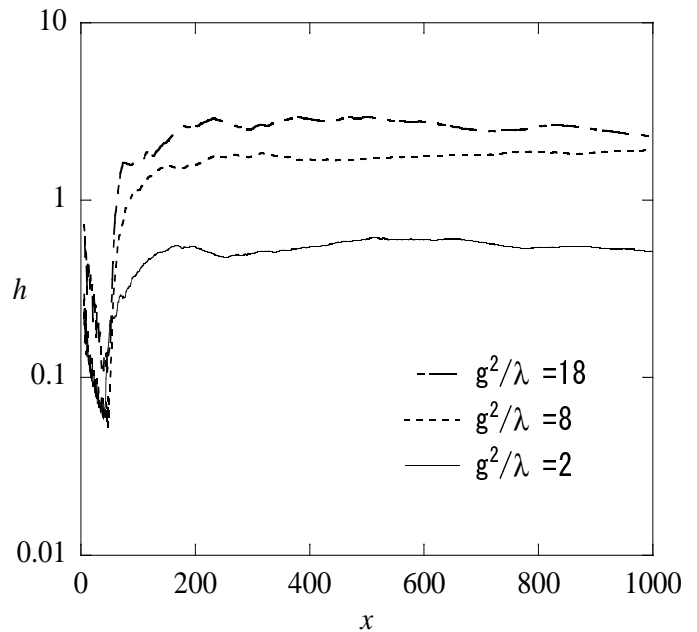}
\caption{Evolution of the maximal Lyapunov exponent $h$
for $g^2/\lambda=2, 8, 18$. We do not implement the backreaction
effect of created particles. We find that the exponent $h$
approaches a constant value in these cases.
}
\label{lyapunov}
\end{center}
\end{figure}

In order to check the existence of chaos, we plot
the evolution of the maximal Lyapunov exponent $h$
for several different values of $g$ in Fig.~\ref{lyapunov}.
The exponent decreases at the initial stage
even though $\chi$ is amplified by parametric resonance.
However $h$ begins to grow after $\chi$
satisfies the Toda-Brumer test (\ref{chin=4}).
The system eventually approaches a phase with
a positive constant $h$, which shows the existence of chaos.
The growth of the maximal Lyapunov exponent
is regarded as a signature of chaos,
since this behavior can not be seen in the absence of chaos.
We checked that $h$ continues to decrease and converges
toward 0 in power if the mixing terms do not exist on the r.h.s. of
Eqs.~(\ref{dphi}) and (\ref{dchi}).

\begin{figure}
\begin{center}
\includegraphics[height=6.0in,width=3.2in]{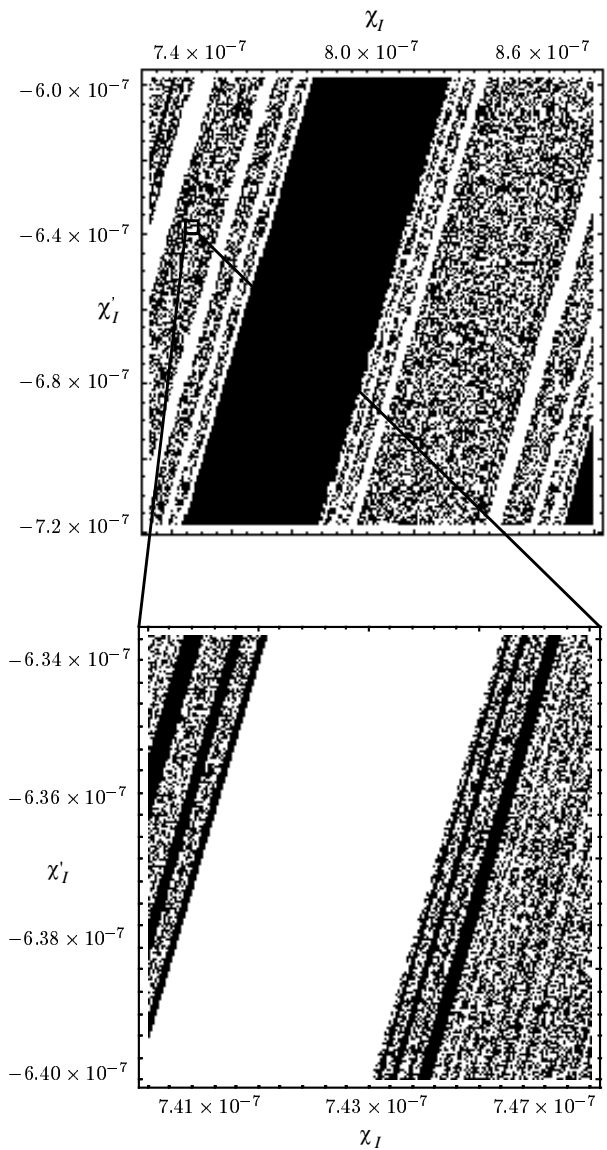}
\caption{A fractal map for $g^2/\lambda=2$.
We show the map of initial conditions $\chi_I$ and
$\dot{\chi}_I$ with exit pockets characterized by
$|\tilde{\chi}| \ge 1.2$. The white color corresponds to
orbits which give $\tilde{\chi} \ge 1.2$, whereas
the black one to orbits which give $\tilde{\chi} \le -1.2$.
The upper panel corresponds to the change of initial conditions
by 0.1\%. The lower panel is an extended figure,
in which initial conditions change by 0.005\%.
These figures exhibit fractal structures, which result
from sensitivity to initial conditions.
}
\label{fractal}
\end{center}
\end{figure}

In Fig.~\ref{fractal} we show a fractal map for $g^2/\lambda=2$
with slight change of initial conditions in terms of $\tilde{\chi}_I$
and $\tilde{\dot{\chi}}_I$. We set exit pockets when the field $\chi$
becomes larger than $|\tilde{\chi}|=1.2$.
When an orbit reaches a pocket we assign colors to many initial 
conditions
as in the following way;
white if an orbit falls down to an upper pocket ($\tilde{\chi}>1.2$),
black if it falls down to a lower pocket ($\tilde{\chi}<-1.2$).
Figure \ref{fractal} is the result of the above manipulation,
which shows that the map of initial conditions is fractal.
This means that orbits are sensitive to initial data,
thereby showing the existence of chaos.

The above discussion neglects the backreaction effect of
created particles. If we account for it as
a Hartree approximation, Eqs.~(\ref{back2}), (\ref{back3}),
(\ref{back}), (\ref{dphi}) and (\ref{dchi}) are modified by replacing
the $\phi^2$, $\chi^2$ and $\phi^3$ terms for
$\phi^2+\langle \delta \phi^2 \rangle$, $\chi^2+\langle \delta \chi^2 
\rangle$
and $\phi^3+3\phi \langle \delta \phi^2 \rangle$, respectively 
\cite{KLS2}.
By Eq.~(\ref{dphi}) the backreaction becomes important when
\begin{eqnarray}
\label{backcon}
\sqrt{\langle \delta \tilde{\chi}^2 \rangle} \gtrsim
\sqrt{\frac{3\lambda}{g^2}}\,\tilde{\phi}\,.
\end{eqnarray}
This is similar to the necessary condition for chaos for
the background field $\chi$, see Eq.~(\ref{chin=4}).
When $g^2/\lambda={\cal O}(1)$, the Toda-Brumer test
is satisfied before the condition (\ref{backcon}) is fulfilled.
As seen in Fig.~\ref{chidelchi} both $\tilde{\chi}^2$ and
$\langle \delta \tilde{\chi}^2 \rangle$
have similar amplitudes initially, but the growth of sub-Hubble 
fluctuations
occurs later than that of $\tilde{\chi}^2$.
At the time when the Toda-Brumer test (\ref{chin=4})
is satisfied ($x \simeq 45$), $\langle \delta \tilde{\chi}^2 \rangle$
is much smaller than $\tilde{\chi}^2$ for $g^2/\lambda=2$.
The backreaction effect becomes important around $x=60$
in this case.

\begin{figure}
\begin{center}
\includegraphics[height=3.0in,width=3.2in]{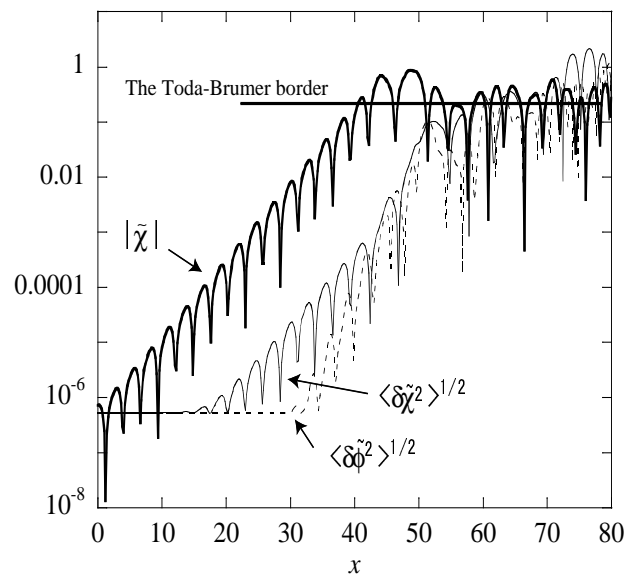}
\caption{Evolution of $\tilde{\chi}$, $\langle \delta
\tilde{\chi}^2\rangle^{1/2}$ and $\langle \delta 
\tilde{\phi}^2\rangle^{1/2}$
(normalized by $m_p$) for $g^2/\lambda=2$
when the backreaction effect is taken into account.
The horizontal line corresponds to the Toda-Brumer border.
}
\label{chidelchiback}
\end{center}
\end{figure}

By implementing the backreaction as a Hartree approximation,
we find that this typically tends to work to suppress the growth
of field fluctuations.
As illustrated in Fig.~\ref{chidelchiback} the fluctuations do not 
exhibit rapid
growth after the backreaction begins to work ($x \gtrsim 60$).
Nevertheless we need to caution that linear perturbation theory
is no longer valid at this stage. For completeness it is required
to account for the mode-mode coupling (rescattering)
between the fluctuations \cite{KT}. In fact we found a
numerical instability for $x \gtrsim 80$ in the simulation of
Fig.~\ref{chidelchiback}, which signals the limitation of
the Hartree approximation.
It is of interest to see the effect of chaos at this fully nonlinear 
stage,
but this is a non-trivial problem because of the complex nature
of reheating. Note that the chaotic period ends at some time
to complete reheating.
It is difficult to judge when chaos ends in our system,
since the mechanism for the decay of $\phi$ and $\chi$
after preheating is not completely known.

\begin{figure}
\begin{center}
\includegraphics[height=3.0in,width=3.2in]{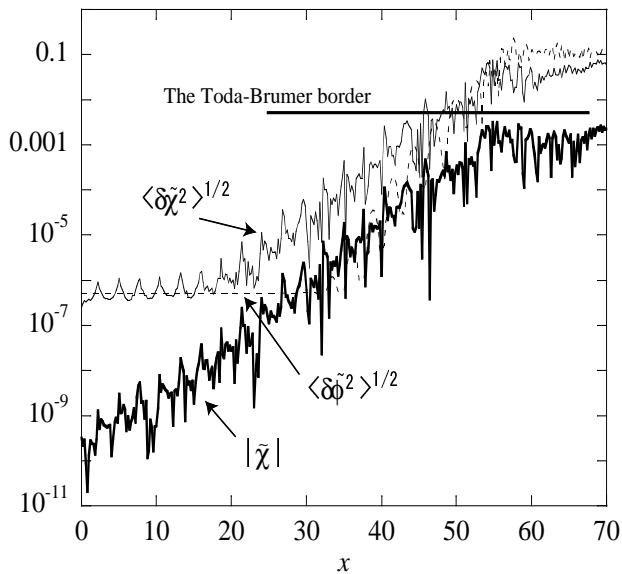}
\caption{As in Fig.~\ref{chidelchiback} for $g^2/\lambda=5000$.
}
\label{g5000}
\end{center}
\end{figure}

When $g^2/\lambda={\cal O}(1)$ one can find out the existence of chaos
during a short period before the backreaction begins to work.
As we already mentioned, the criterion for chaos is
given by Eq.~(\ref{chin=4}), whereas the criterion for the
backreaction corresponds to Eq.~(\ref{backcon}).
The initial value of the quasi-homogeneous field $\chi$
gets smaller for larger $g^2/\lambda$, as illustrated in 
Fig.~\ref{n4}.
This means that the condition (\ref{backcon}) tends to be satisfied
prior to the time at which the quasi-homogeneous field $\chi$
grows to satisfy the Toda-Brumer test (\ref{chin=4}).
In Fig.~\ref{g5000} we plot the evolution of the system for
$g^2/\lambda=5000$ with the backreaction effect of
created particles.
In this case the backreaction becomes important around $x=50$
before the quasi-homogeneous field $\chi$ increases sufficiently
to satisfy the Toda-Brumer test.
Therefore, when $g^2/\lambda \gg 1$, we do not find a signature
of chaos before the perturbations reach a nonlinear regime.

For the coupling $g$ that belongs to the resonance bands
$n(2n-1)<g^2/\lambda<n(2n+1)$, we find that 
chaos occurs for
\begin{eqnarray}
g^2/\lambda < {\cal O}(10)\,,
\end{eqnarray}
before the backreaction begins to work.
We note that this result is obtained by using the value
$\sqrt{\langle \delta \chi^2_{k} (t_{f}) \rangle_{k<k_f}}$ derived 
in Sec.~II  as the initial condition of $\chi$ at the beginning of 
preheating.
When $g^2/\lambda > {\cal O}(10)$ the field fluctuation
$\langle \delta \tilde{\chi}^2 \rangle$ satisfies the condition
(\ref{backcon}) prior to the time at which the necessary condition
for chaos is fulfilled. 
Provided that $g^2/\lambda \gg 1$, the standard
Floquet theory of parametric resonance \cite{Kaiser,GKLS} is valid
at the linear level, since the effect of the quasi-homogeneous field 
$\chi$
is not important relative to its perturbations on sub-Hubble scales.

\section{Quadratic potential}

For the quadratic potential the system can not be
reduced to the analysis in Minkowski spacetime
by introducing conformal quantities.
Therefore the analysis in the quadratic potential is more complicated 
than
in the case of the quartic potential
in an expanding background.

We start our analysis by studying the two-field dynamics in a
frictionless background.
In this case the fields oscillate coherently without
an adiabatic damping due to cosmic expansion.
The system has the field equations
corresponding to $H=0$ in Eqs.~(\ref{back2}) and  (\ref{back3})
together with the constraint
\begin{eqnarray}
\label{energy}
E=\frac12 \dot{\phi}^2+\frac12 \dot{\chi}^2+
V(\phi, \chi)\,,
\end{eqnarray}
where $E$ is conserved.
Unlike the case of an expanding background, the field $\chi$ is
enhanced only when the system is inside of resonance bands
from the beginning of preheating.
This comes from the fact that the field $\chi$ does not shift to other
stability/instability bands in the absence of cosmic
expansion \cite{KLS2}. We wish to study the existence of chaos
for the coupling $g$ that leads to parametric excitation of
$\chi$ in an expanding background ($g \gtrsim 3.0 \times 10^{-4}$).
First we carry out the analysis in a conserved Hamiltonian system
given above and then proceed to the case in which the expansion
of universe is taken into account.

As illustrated in Fig.~\ref{n2}, the quasi-homogeneous $\chi$
field at the end of inflation is estimated to be
$\chi_f \lesssim 10^{-9}$-$10^{-8}m_{p}$ for the coupling
$g \gtrsim 3.0 \times 10^{-4}$, which is smaller than that
in the quartic potential with $g^2/\lambda={\cal O}(1)$.
Hence the field $\chi$ for the quadratic potential
is more strongly suppressed during inflation
for the values of $g$ relevant to efficient preheating.

\begin{figure}
\begin{center}
\includegraphics[height=3.0in,width=3.2in]{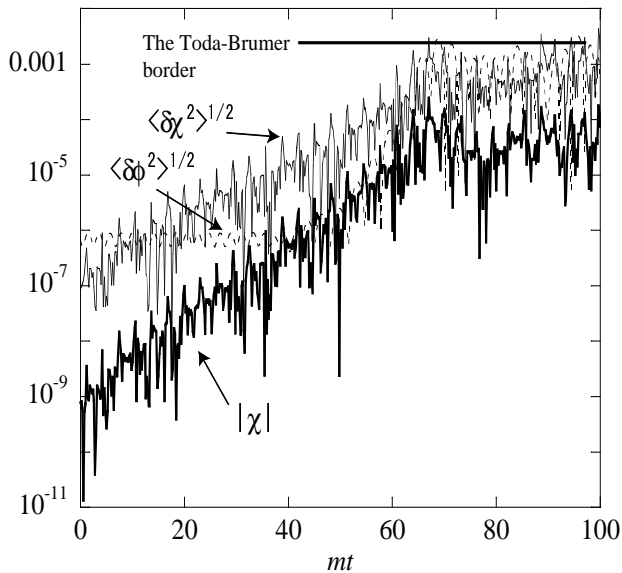}
\caption{Evolution of $\chi$, $\langle \delta \chi ^2\rangle^{1/2}$ 
and
$\langle \delta \phi^2\rangle^{1/2}$ (normalized by $m_p$)
for $g=3.0\times 10^{-4}$
without the friction due to cosmic expansion.
The horizontal line represents the border of the Toda-Brumer test.
We implement the backreaction effect of created particles
as a Hartree approximation.
}
\label{qua1}
\end{center}
\end{figure}

In Fig.~\ref{qua1} we plot the evolution of the background field 
$\chi$
together with $\sqrt{\langle \delta \chi ^2\rangle}$ and
$\sqrt{\langle \delta \phi^2\rangle}$ for the coupling $g=3.0\times 
10^{-4}$.
Note that we implement the backreaction of sub-Hubble
field fluctuations as a Hartree approximation.
In this case the Toda-Brumer test gives the condition
$\chi/m_p>1.9\times 10^{-3}$ by Eq.~(\ref{chin=2}).
For the quadratic potential the backreaction begins to work for
$\langle \delta \chi^2 \rangle \gtrsim m^2/g^2$,
which is basically a similar condition to the Toda-Brumer test
for the background field $\chi$.
As shown in Fig.~\ref{qua1} the quasi-homogeneous field $\chi$
does not satisfy the Toda-Brumer test, since the backreaction
becomes important before $\chi$ grows sufficiently.

\begin{figure}
\begin{center}
\includegraphics[height=3.0in,width=3.2in]{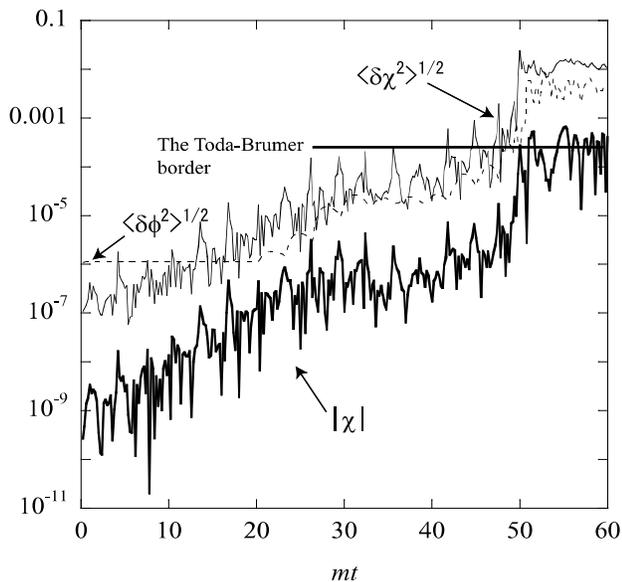}
\caption{As in Fig.~\ref{qua1} with $g=3.0 \times 10^{-3}$.
}
\label{qua2}
\end{center}
\end{figure}

The quasi-homogeneous $\chi$ field at the end of inflation gets 
smaller for larger $g$, see Eq.~(\ref{dchik}).
Figure \ref{qua2} shows the evolution of the system
for $g=3.0 \times 10^{-3}$, in which case the initial variance of the
field $\chi$ is suppressed relative to the case $g=3.0 \times 
10^{-4}$.
Although the condition for chaos using
the Toda-Brumer test (\ref{chin=2}) gets milder for larger $g$,
this property is compensated by
the suppression of the quasi-homogeneous field $\chi$
at the beginning of preheating.
Therefore it is difficult to satisfy the necessary condition for chaos
before the backreaction begins to work.
We carried out numerical simulations for other values of
$g$ and found that the signature of chaos is not seen
in the frictionless system as long as the backreaction effect
is taken into account.

If we implement the effect of cosmic expansion, the energy of the
system given by Eq.~(\ref{energy}) decreases.
In fact the time-derivative of $E$ is given by
\begin{eqnarray}
\frac{{\rm d}E}{{\rm d}t}=-3H(\dot{\phi}^2+\dot{\chi}^2)
\simeq -3HE\,,
\end{eqnarray}
where we used the approximation $E \simeq \dot{\phi}^2+\dot{\chi}^2$.
Then the energy lost during one oscillation of
the inflaton ($\Delta t \simeq 1/m$) is estimated as
\begin{eqnarray}
\frac{\Delta E}{E} \simeq -3\frac{H}{m} \simeq
-\sqrt{24\pi} \frac{\phi}{m_p}\,.
\end{eqnarray}
Therefore the energy loss is large at the beginning of
preheating ($|\Delta E/E| \gtrsim 0.1$),
but it becomes smaller and smaller with time.
The analysis in the frictionless system can be used in an expanding
background when the condition, $|\Delta E/E| \ll 1$, is satisfied.

In addition to the energy loss, we need to
caution that the structure of
resonance changes in the presence of cosmic expansion.
The field $\chi$ passes many instability/stability bands, which is
generally called {\it stochastic resonance} \cite{KLS2}.
Parametric resonance ends when the resonance parameter,
$q=g^2\phi^2/(4m^2)$, drops down to less than of order unity,
whose property is different from the analysis in Minkowski spacetime.

In spite of above complexities, it is possible to check whether
the signature of chaos is seen or not in a linear perturbation regime.
Note that the Toda-Brumer condition is valid in an expanding
background. We run our numerical code including
the backreaction of sub-Hubble field fluctuations for the coupling
$g$ relevant to efficient preheating ($g \gtrsim 3.0 \times 10^{-4}$)
and find that the background $\chi$ stops growing by the backreaction
effect before the Toda-Brumer condition is satisfied.
This property is similar to the case discussed in the frictionless
background. Therefore chaos does not appear for the quadratic
potential at least in a regime before the backreaction sets in.

\section{Conclusions}

In this paper we discussed chaotic dynamics in two-field preheating
with monomial inflaton potentials $V(\phi)=V_0\phi^n$.
A scalar field $\chi$ coupled to inflaton with an interaction
$(1/2)g^2\phi^2\chi^2$ is amplified by parametric resonance
when the inflaton oscillates coherently.
As long as the background field $\chi$ is not
strongly damped in an inflationary epoch, $\chi$ can grow to
the same order as $\phi$, thereby giving rise to a possibility
of chaos during preheating.

First we estimated the amplitude of the quasi-homogeneous
$\chi$ field at the beginning of reheating by considering 
the variance of the $\chi$ fluctuations on scales larger than 
the Hubble radius at the end of inflation.
The variance $\sqrt{\langle \delta \chi^2_{k}
(t_{f}) \rangle_{k<k_f}}$ is plotted as a function of the coupling 
$g$ in Figs.~\ref{n2} and \ref{n4}.
For the quartic inflaton potential with $g^2/\lambda={\cal O}(1)$
there exist large-scale perturbation modes ($k<k_{c}$) 
which are not strongly suppressed during inflation.
Then this gives large contribution to the total variance of 
$\chi$ relative to the case $g^2/\lambda \gg 1$ at the end of 
inflation.

Typically the background field $\chi$ is assumed to be
negligible in standard analysis of particle creations in
preheating, but its presence leads to a mixing between two fields.
This gives rise to a chaotic instability in addition to the
enhancement of perturbations by parametric resonance. We note that
standard Floquet theory using Mathieu or Lame equation ceases to
be valid when the chaos is present. This new channel of
instability can alter the maximum size of field fluctuations if
two dynamical fields do not decay for a long time.

In order to study whether chaos really appears or not, we solved the
background equations (\ref{back2}), (\ref{back3}) and (\ref{back})
together with perturbed equations (\ref{dphi}) and (\ref{dchi}).
For a quartic potential ($n=4$ and $V_0=\lambda/4$)
parametric resonance can occur for the coupling $g^2/\lambda$ of order
unity. In this case the quasi-homogeneous field $\chi$ satisfies the 
Toda-Brumer test (\ref{chin=4}) 
before the backreaction effect of created particles
becomes important. Since this is only the necessary condition for 
chaos, we also
evaluated Lyapunov exponents for $g^2/\lambda={\cal O}(1)$ in order 
to confirm the
presence of chaos. We find that the maximal Lyapunov exponent begins 
to increase
toward a positive constant value after the Toda-Brumer condition is 
satisfied,
which shows the existence of chaos.
Our analysis using a Fractal map also implies that the system exhibits
chaotic behavior when $g^2/\lambda$ is not too much larger than unity.
For larger $g^2/\lambda$, the backreaction of field fluctuations on 
sub-Hubble
scales works earlier than the time at which the Toda-Brumer test is
satisfied for the background field $\chi$. We find signatures of chaos
for $g^2/\lambda <{\cal O} (10)$ at the linear regime before the 
backreaction
begins to work. For $g^2/\lambda \gg 1$ the quasi-homogeneous field 
$\chi$ gets smaller
at the beginning of preheating, in which case perturbations
enter a nonlinear region before chaos can be seen.

The system with a quadratic ($n=2$) potential can not
be effectively reduced to the analysis in Minkowski spacetime
unlike the quartic potential.
We first analyze the preheating dynamics in a frictionless
background and then proceed to the case in which the
expansion of universe is taken into account.
In this model the background field $\chi$
does not grow sufficiently to satisfy  the Toda-Brumer test
in the presence of the backreaction effect of created particles.
We find that this result holds both in Minkowski and expanding
backgrounds. Therefore chaos can not be observed at least
at the linear stage of preheating before the backreaction sets in.

When chaos is present, this means that the second field $\chi$
gives a non-negligible contribution for the background dynamics.
This generally gives a strong correlation between adiabatic and
isocurvature metric perturbations \cite{corre}, which can lead to
the amplification of curvature perturbations for the self-coupling
inflation model \cite{zetagrow}. This reduces the tensor to scalar
ratio $r$, which is favorable from the observational point of
view. Of course the ratio $r$ is not sufficient to judge whether
the model is rescued or not, since large contribution of
isocurvature perturbations modifies the CMB power spectrum. It is
still interesting that the appearance of chaos has a possibility
to reduce the ratio $r$.

In this work we did not study the nonlinear dynamics of the system
during which the mode-mode coupling between perturbations plays an
important role. We expect that the presence of chaos persists even
in such a nonlinear stage provided that two interacting fields
$\phi$ and $\chi$ are dynamically important. The chaos would
finally disappear after the energy densities of scalar fields are
converted to that of radiation. It is of interest to extend our
analysis to such a regime including Born decays of scalar fields
for a complete understanding of chaos in reheating.

\section*{ACKNOWLEDGMENTS}
It is a pleasure to thank Kei-ichi Maeda and Alexei Starobinsky for 
useful discussions.
This work was partially supported by a Grant for The 21st Century COE 
Program
(Holistic Research and Education Center for Physics Self-organization 
Systems)
at Waseda University.


\end{document}